\documentclass{article}%
\usepackage{amsfonts}
\usepackage{amsmath}
\usepackage{cite}
\usepackage{amssymb}
\usepackage{charter}
\usepackage{graphicx}%
\providecommand{\U}[1]{\protect \rule{.1in}{.1in}}
\makeatletter
\def \@cite#1#2{\textsuperscript{[{#1\if@tempswa , #2\fi}]}}
\makeatother

\begin{document}

\title{A new Coherent-Entangled state generated by an asymmetric beam splitter and
its applications}
\author{Xu Ma, Shuang-Xi Zhang\thanks{\emph{Corresponding author}:
shuangxi@mail.ustc.edu.cn} and Gang Ren\\Department of Material
Science and Engineering, University of \\Science and Technology of
China, Hefei, Anhui 230026, China} \maketitle

\begin{abstract}
A new kind of tripartite non-symmetric coordinate coherent-entangled state
(TNCCES) $\left \vert \beta,\gamma,x\right \rangle $ is proposed which exhibits
the properties of both coherence and entanglement and makes up a new quantum
mechanical representation.We investigate some properties of TNCCES such as
completeness and orthogonality which prove it is just a tripartite complete
continuous coordinate base. A protocol for generating TNCCES is proposed using
asymmetric beam splitter. And in application of TNCCES, we find its
corresponding Wigner operator and carry out its marginal distribution form;
further a new tripartite entangled squeezed operator is also presented. The
multipartite CES and its generation are also disussed.



\end{abstract}

\section{Introduction}

Quantum entangled states have been vastly studied by physicists due to their
potential uses in quantum information and quantum communication. The concept
of entanglement, originated by Einstein, Podolsky and Rosen (EPR) for arguing
the incompleteness of quantum mechanics, plays a key role in understanding
some fundamental problems in quantum mechanics and quantum optics. In a
quantum entangled system, a measurement performed on one part of the system
provides information about the remaining part, and this is now known as the
basic feature of quantum mechanics, weird though it seems. For a good
understanding entanglement, Re[1] will be useful. Beyond all entangled states,
the continuous variable entangled states are of great application in quantum
optics and atomics area, where the continuous variables are just the
quadrature phase of optics field. Detail acquaintance of continuous variable
can refer to Ref.[2,3]. And it may be infered that continuous variable
entanglement such as topological entanglement may play an important roll in
understanding famous and obscure phenomenons in low temperature physics such
as the fractional electron charge effect\cite{4}. On the other hand, the
theoretical research has go ahead of experiment to construct various
continuous variable entangled states :idealized EPR state $\left \vert
\eta \right \rangle $\cite{5}, two mode coherent entangled state $\left \vert
\alpha,x\right \rangle $\cite{6}, and arbitrary multi-mode entangled state
\cite{7} and so on. All these mentioned states are constructed and
property-analyzed basing on IWOP technique\cite{8,9,10,11}. Construst to
classical quantum optical states, these states present nonclassical properties
such as partial non-positive of Wigner distributive and Mandel factor\cite{12}%
, divergence in special point in phase space of Glauber-Shudasan
representation\cite{13}. Among these states coherent-entangled state (CES) is
of special interesting due to its intrinsic nature of the merging of coherence
and entanglement. As far as we know, multi-mode of coordinate symmetric
coherent-entangled state (CCES) have been down in Ref.[6,15], but the
non-symmetric CCES haven't done yet, and we find it not trivial to generalize
symmetric to non-symmetric ones. Although these problems ahead, we will
construct definite expression of three-mode non-symmetric coherent-entangled
state $\left \vert \beta,\gamma,x\right \rangle $. As coherent entangled state,
one of its important roll it can play is as the continuous base representation
in Hilbert space of square-integrated property. So we will check its
completeness and orthogonality. Further we investigate its conjugate state,
and basing on its non-symmetry, the non-symmetric Wigner operator of
coordinate eigenvalue $x$ is constructed,and its marginal distribution
operator after integrated out one quadrature phase is gotten. As one
application of $\left \vert \beta,\gamma,x\right \rangle $, the squeezed
operator basing on IWOP technique is constructed.

This paper is arranged as follows, firstly in Sec. 2 the explicit form of
tripartite CES $\left \vert \beta,\gamma,x\right \rangle $ in Fock space by
virtue of the technique of integration within an ordered product of operators
(IWOP), and its main properties are analysed in Sec. 3. In Sec. 4, we discuss
how to generate the ideal $\left \vert \beta,\gamma,x\right \rangle $ state as
the output of an asymmetric beam splitter. In Sec. 5 we discuss briefly for
some probably applications of $\left \vert \beta,\gamma,x\right \rangle $ in
quantum optics. In Sec.6, we introduce the new multipartite CES and its generation.

\section{The Introduction of Triparticle non-symmetric Coordinate
Coherent-Entangled State (TNSCE)}

For a physical state one hopes that it can span a complete space. For example,
the Fock state and the coherent state are both (over) complete. It has been
shown\cite{7} that by constructing miscellaneous normally ordered Gaussian
integration operators, which are unity operators, and then considering their
decomposition of unity we may derive new quantum mechanical states possessing
the completeness relation and orthogonality. For coherent state, using the
normal ordering form of vacuum state projector\cite{16,17} $\left \vert
0\rangle \langle0\right \vert =:\exp(-a^{\dag}a):$, where $:\ :$ is a symbol of
normal ordering, making use of following formula
\begin{equation}
\int \frac{d^{2}z}{\pi}\left.  :\exp \left[  -(z^{\ast}-a^{\dag})(z-a)\right]
:\right.  =1. \label{1}%
\end{equation}
After making the decomposition
\begin{equation}
:\exp \left[  -(z^{\ast}-a^{\dag})(z-a)\right]  :=\left \vert z\rangle \langle
z\right \vert \label{2}%
\end{equation}
the form of coherent state $\left \vert z\right \rangle =\exp(-\frac{|z|^{2}}%
{2}+za^{\dag})\left \vert 0\right \rangle $ emerges.

Similarly, by examining
\begin{equation}
\int \frac{d^{2}\eta}{\pi}\left.  :\exp \left[  -(\eta^{\ast}-a_{1}^{\dag}%
+a_{2})(\eta-a_{1}+a_{2}^{\dag})\right]  :\right.  =1 \label{3}%
\end{equation}
and decomposing the integrand in Eq.(\ref{3}), we observe the emergence of
biparticle ideal EPR state $\left \vert \eta \right \rangle $\cite{5}.

Noticing $[(\mu X_{1}+\nu X_{2}),(\nu a_{1}-\mu a_{2})]=0$, we can get the
expression of bipartite coherent-entangled state\cite{18} in this way as
\begin{equation}
\left \vert \alpha,x\right \rangle =\exp \left[  -\frac{1}{2}x^{2}-\frac{1}%
{4}|\nu \alpha|^{2}+\lambda \alpha a_{1}^{\dag}+\frac{2x-\alpha \mu}{2\lambda
}\left(  \mu a_{1}^{\dag}+\nu a_{2}^{\dag}\right)  -\frac{1}{4\lambda^{2}%
}\left(  \mu a_{1}^{\dag}+\nu a_{2}^{\dag}\right)  ^{2}\right]  \left \vert
00\right \rangle ,\label{4}%
\end{equation}
where $\mu^{2}+\nu^{2}=2\lambda^{2}$ to make sure it can be prepared by
beam-splitter. Using the bosonic communicative relation $[a_{i},a_{j}^{\dag
}]=\delta_{ij}$, we have
\begin{align}
a_{1}\left \vert \alpha,x\right \rangle  &  =\left[  \alpha \lambda+\frac{\mu
}{\lambda}\left(  x-\frac{1}{2}\alpha \mu \right)  -\frac{\mu}{2\lambda^{2}%
}\left(  \mu a_{1}^{\dag}+va_{2}^{\dag}\right)  \right]  \left \vert
\alpha,x\right \rangle ,\nonumber \\
a_{2}\left \vert \alpha,x\right \rangle  &  =\left[  \frac{v}{\lambda}\left(
x-\frac{1}{2}\alpha \mu \right)  -\frac{v}{2\lambda^{2}}\left(  \mu a_{1}^{\dag
}+va_{2}^{\dag}\right)  \right]  \left \vert \alpha,x\right \rangle ,
\end{align}
which satisfy the following eigenequations
\begin{equation}
\frac{1}{2}\left(  \mu X_{1}+vX_{2}\right)  \left \vert \alpha,x\right \rangle
=\frac{\lambda x}{\sqrt{2}}\left \vert \alpha,x\right \rangle ,\quad \left(
va_{1}-\mu a_{2}\right)  \left \vert \alpha,x\right \rangle =va\lambda \left \vert
\alpha,x\right \rangle ,
\end{equation}
which means $\left \vert \alpha,x\right \rangle $is actually the common
eigenvector of $(\mu X_{1}+\nu X_{2})$ and $(\nu a_{1}-\mu a_{2})$.

However, if we want to obtain the expression of TNCCES, it may become tedious
to construct such a complex quadratic gaussian polynomial of three-mode of
generate operator in entangled form to derive TNCCES. Fortunately, we can
stride over this problem just oppositely, first construct the tripartite
counterpart formally analogue to bipartite CCES, then check it satisfies the
similar relationship as Eq.(\ref{3}). For $[(\mu X_{1}+\nu X_{2}+\tau
X_{3}),(\nu a_{1}-\mu a_{2})]=[(\mu X_{1}+\nu X_{2}+\tau X_{3}),(\tau
a_{2}-\nu a_{3})]=0$, $\left \vert \beta,\gamma,x\right \rangle $ can be
introduced as%
\begin{align}
\left \vert \beta,\gamma,x\right \rangle  &  =\exp \left \{  \left[  -\frac{3}%
{4}x^{2}-\frac{1}{6\nu}(\beta^{\ast}\gamma+\beta \gamma^{\ast})\mu \tau
^{2}-\frac{1}{6}|\gamma|^{2}\tau^{2}\left(  1+\frac{\mu^{2}}{\nu^{2}}\right)
-\frac{1}{6}|\beta|^{2}\left(  \nu^{2}+\tau^{2}\right)  \right]  \right.
\nonumber \\
&  +\frac{1}{3\lambda}\left[  \left(  \beta(\nu^{2}+\tau^{2})+\frac{\gamma
\mu \tau^{2}}{\nu}+3x\mu \right)  a_{1}^{\dag}+\left(  -\beta \mu \nu+\gamma
\tau^{2}+3x\nu \right)  a_{2}^{\dag}+\quad \left(  -\frac{\gamma(\mu^{2}+\nu
^{2})\tau}{\nu}-\beta \mu \tau+3x\tau \right)  a_{3}^{\dag}\right] \nonumber \\
&  \quad \left.  -\frac{1}{6\lambda^{2}}\left(  \mu a_{1}^{\dag}+\nu
a_{2}^{\dag}+\tau a_{3}^{\dag}\right)  ^{2}\right \}  \left \vert
000\right \rangle , \label{7}%
\end{align}
where $\lambda^{2}=\frac{1}{3}\left(  \mu^{2}+\nu^{2}+\tau^{2}\right)  $ as
that done for bipartite to make sure it can be generated by beam-splitter
which will be instructed in Sec. 4.

Acting $a_{i}(i=1,2,3)$ on $\left \vert \beta,\gamma,x\right \rangle $, we have
\begin{align}
a_{1}\left \vert \beta,\gamma,x\right \rangle  &  =\frac{1}{3\lambda}\left[
\left(  \beta \left(  v^{2}+\tau^{2}\right)  +\frac{\gamma \mu \tau^{2}}{v}%
+3x\mu \right)  -\frac{\mu}{\lambda}\left(  \mu a_{1}^{\dag}+va_{2}^{\dag}+\tau
a_{3}^{\dag}\right)  \right]  \left \vert \beta,\gamma,x\right \rangle
,\nonumber \\
a_{2}\left \vert \beta,\gamma,x\right \rangle  &  =\frac{1}{3\lambda}\left[
\left(  -\beta \mu \nu+\gamma \tau^{2}+3x\upsilon \right)  -\frac{v}{\lambda
}\left(  \mu a_{1}^{\dag}+va_{2}^{\dag}+\tau a_{3}^{\dag}\right)  \right]
\left \vert \beta,\gamma,x\right \rangle ,\nonumber \\
a_{3}\left \vert \beta,\gamma,x\right \rangle  &  =\frac{1}{3\lambda}\left[
\left(  -\frac{\gamma \tau \left(  \mu^{2}+v^{2}\right)  }{v}-\beta \mu
\tau+3x\tau \right)  -\frac{\tau}{\lambda}\left(  \mu a_{1}^{\dag}+va_{2}%
^{\dag}+\tau a_{3}^{\dag}\right)  \right]  \left \vert \beta,\gamma
,x\right \rangle . \label{9}%
\end{align}

From the Eq.(\ref{9}) , we obtain the eigenequations of $\left \vert
\beta,\gamma,x\right \rangle $%
\begin{align}
\frac{1}{3}\left(  \mu X_{1}+vX_{2}+\tau X_{3}\right)  \left \vert \beta
,\gamma,x\right \rangle  &  =\frac{\lambda x}{\sqrt{2}}\left \vert \beta
,\gamma,x\right \rangle ,\nonumber \\
\left(  va_{1}-\mu a_{2}\right)  \left \vert \beta,\gamma,x\right \rangle  &
=\nu \beta \lambda \left \vert \beta,\gamma,x\right \rangle ,\nonumber \\
\left(  \tau a_{2}-va_{3}\right)  \left \vert \beta,\gamma,x\right \rangle  &
=\tau \gamma \lambda \left \vert \beta,\gamma,x\right \rangle . \label{10}%
\end{align}

So we see that $\left \vert \beta,\gamma,x\right \rangle $ is actually the
common eigenvector of $\frac{1}{3}(\mu X_{1}+\nu X_{2}+\tau X_{3})$, $(\nu
a_{1}-\mu a_{2})$ and $(\tau a_{2}-\nu a_{3})$. Eq.(\ref{10}) just corresponds
to the eigen-relationships of bipartite example.

\section{Main Properties Of $\left \vert \beta,\gamma,x\right \rangle $}

In Sec.2, we construct the TSSCES $\left \vert \beta,\gamma,x\right \rangle $
just oppositely to traditional ways, and in this section we will check its
orthogonality and completeness, to prove it span the whole Hilbert space which
is basic property for a kind of representation base. And as the eigenstates of
non-symmetric sum of coordinates of three modes, we can anticipate its
analogue to single mode coordinate eigen-state, conjugate state as well.
Further more what can't be forgotten is that it presents the entanglement
between coordinates of three mode and eigenrelationship of various linear
combination of 1,2,3. Basing on these special properties of $\left \vert
\beta,\gamma,x\right \rangle $, we may construct dynamics of tripartite
entangled state driven by system and environment Hamilton which will not be
referred in this article. what is the focus of this article is the basic
properties of $\left \vert \beta,\gamma,x\right \rangle $.

\subsection{Orthogonal Property}

We investigate whether $\left \vert \beta,\gamma,x\right \rangle $ is mutual
orthogonal or not. Explicitly, using the eigen equations (\ref{10}), we
examine the following matrix elements:%
\begin{equation}
\left \langle \beta^{\prime},\gamma^{\prime},x^{\prime}\right \vert \frac{\mu
X_{1}+\nu X_{2}+\tau X_{3}}{3}\left \vert \beta,\gamma,x\right \rangle
=\frac{\lambda x^{\prime}}{\sqrt{2}}\left \langle \beta^{\prime},\gamma
^{\prime},x^{\prime}|\beta,\gamma,x\right \rangle =\frac{\lambda x}{\sqrt{2}%
}\left \langle \beta^{\prime},\gamma^{\prime},x^{\prime}|\beta,\gamma
,x\right \rangle
\end{equation}
which leads to%
\begin{equation}
\left \langle \beta^{\prime},\gamma^{\prime},x^{\prime}|\beta,\gamma
,x\right \rangle =\delta(x^{\prime}-x)
\end{equation}
To derive the exact express of $\left \langle \beta^{\prime},\gamma^{\prime
},x^{\prime}|\beta,\gamma,x\right \rangle $, we will use the over-completeness
relation of the three-mode coherent state
\begin{equation}
\int \frac{d^{2}z_{1}d^{2}z_{2}d^{2}z_{3}}{\pi^{3}}\left \vert z_{1},z_{2}%
,z_{3}\rangle \langle z_{1},z_{2},z_{3}\right \vert =1
\end{equation}
where
\begin{align}
\left \vert z_{1},z_{2},z_{3}\right \rangle  &  =D_{1}(z_{1})D_{2}(z_{2}%
)D_{3}(z_{3})\left \vert 000\right \rangle \nonumber \\
&  =\exp \left[  -\frac{1}{2}\left(  |z_{1}|^{2}+|z_{2}|^{2}+|z_{3}%
|^{2}\right)  +z_{1}a_{1}^{\dag}+z_{2}a_{2}^{\dag}+z_{3}a_{3}^{\dag}\right]
\left \vert 000\right \rangle
\end{align}
and $D_{i}(z)=\exp(za_{i}^{\dag}-z^{\ast}a_{i})$.

Using the definition of Tripartite CES in Eq.(\ref{9}), the overlap is%
\begin{align}
\left \langle z_{1},z_{2},z_{3}|\beta,\gamma,x\right \rangle  &  =\exp \left \{
\left[  -\frac{3}{4}x^{2}-\frac{1}{6\nu}(\beta^{\ast}\gamma+\beta \gamma^{\ast
})\mu \tau^{2}-\frac{1}{6}|\gamma|^{2}\tau^{2}\left(  1+\frac{\mu^{2}}{\nu^{2}%
}\right)  -\frac{1}{6}|\beta|^{2}\left(  \nu^{2}+\tau^{2}\right)  \right]
\right. \nonumber \\
&  +\frac{1}{3\lambda}\left[  \left(  \beta(\mu^{2}+\tau^{2})+\frac{\gamma
\mu \tau^{2}}{\nu}+3x\mu \right)  z_{1}^{\ast}+\left(  -\beta \mu \nu+\gamma
\tau^{2}+3x\nu \right)  z_{2}^{\ast}+\left(  -\frac{\gamma(\mu^{2}+\nu^{2}%
)\tau}{\nu}-\beta \mu \tau+3x\tau \right)  z_{3}^{\ast}\right] \nonumber \\
&  \left.  -\frac{1}{6\lambda^{2}}\left(  \mu z_{1}^{\ast}+\nu z_{2}^{\ast
}+\tau z_{3}^{\ast}\right)  ^{2}-\frac{1}{2}\left(  |z_{1}|^{2}+|z_{2}%
|^{2}+|z_{3}|^{2}\right)  \right \}  .
\end{align}
\quad \ To calculate
\begin{align}
&  \left \langle \beta^{\prime},\gamma^{\prime},x^{\prime}|\beta,\gamma
,x\right \rangle =\int \frac{d^{2}z_{1}d^{2}z_{2}d^{2}z_{3}}{\pi^{3}%
}\left \langle \beta^{\prime},\gamma^{\prime},x^{\prime}|z_{1},z_{2}%
,z_{3}\right \rangle \left \langle z_{1},z_{2},z_{3}|\beta,\gamma,x\right \rangle
\nonumber \\
&  =\exp \left \{  -\frac{\mu^{2}+\nu^{2}}{6\nu^{2}}\left[  \nu^{2}(|\beta
|^{2}+|\beta^{\prime2})+\tau^{2}\left(  |\gamma|^{2}+|\gamma^{\prime2}\right)
\right]  -\frac{\mu}{6\nu}\tau^{2}\left[  \beta \gamma^{\ast}+\beta^{\ast
}\gamma+\beta^{\prime}\gamma^{\prime \ast}+\beta^{\prime \ast}\gamma^{\prime
}-2(\beta \gamma^{\prime \ast}+\beta^{\prime \ast}\gamma)\right]  \right.
\nonumber \\
&  \qquad \qquad+\left.  \frac{\nu^{2}+\tau^{2}}{3\nu^{2}}(\nu^{2}\beta
\beta^{\prime \ast}+\mu^{2}\gamma \gamma^{\prime \ast})\right \}  \delta
(x-x^{\prime}). \label{11}%
\end{align}
In deriving the Eq.(\ref{11}), we have used the mathematical formula
\begin{equation}
\int \frac{d^{2}z}{\pi}\exp \left(  \zeta|z|^{2}+\xi z+\eta z^{\ast}%
+fz^{2}+gz^{\ast2}\right)  =\frac{1}{\sqrt{\zeta^{2}-4fg}}\exp \left[
\frac{-\zeta \xi \eta+\xi^{2}g+\eta^{2}f}{\zeta^{2}-4fg}\right]  ,
\end{equation}
with its convergent condition%
\[
Re(\xi+f+g)<0,\qquad Re\left(  \frac{\zeta^{2}-4fg}{\xi+f+g}\right)  <0,
\]
or%
\[
Re(\xi-f-g)<0,\qquad Re\left(  \frac{\zeta^{2}-4fg}{\xi-f-g}\right)  <0.
\]
and the limiting form of Dirac's delta function
\begin{equation}
\delta(x)=\lim_{\varepsilon \rightarrow{}0}\frac{1}{\sqrt{\pi \varepsilon}}%
\exp \left(  -\frac{x^{2}}{\varepsilon}\right)
\end{equation}

\subsection{Completeness Relation}

Now we shall check whether $\left \vert \beta,\gamma,x\right \rangle $ possesses
the completeness relation. By virtue of the technique of IWOP, and the normal
ordered product of the three-mode vacuum projector
\begin{equation}
\left \vert 000\rangle \langle000\right \vert =\left.  :\exp(a_{1}^{\dag}%
a_{1}+a_{2}^{\dag}a_{2}+a_{3}^{\dag}a_{3}):\right.  ,
\end{equation}
we can smoothly prove the completeness relation of $\left \vert \beta
,\gamma,x\right \rangle $
\begin{align}
&  \tau^{2}\lambda^{2}\iint \frac{d^{2}\beta}{\pi}\frac{d^{2}\gamma}{\pi}%
\int_{-\infty}^{\infty}\frac{dx}{\sqrt{6\pi}}\left \vert \beta,\gamma
,x\right \rangle \left \langle \beta,\gamma,x\right \vert \nonumber \\
&  =\tau^{2}\lambda^{2}\iint \frac{d^{2}\beta}{\pi}\frac{d^{2}\gamma}{\pi}%
\int_{-\infty}^{\infty}\frac{dx}{\sqrt{6\pi}}:\exp \left \{  -\frac{1}{3}\left(
\frac{3}{\sqrt{2}}x-\frac{\mu X_{1}+\nu X_{2}+\tau X_{3}}{\lambda}\right)
^{2}\right. \nonumber \\
&  \qquad-\frac{1}{3}\left[  \left(  \nu \beta^{\ast}-\frac{\nu a_{1}^{\dag
}-\mu a_{2}^{\dag}}{\lambda}\right)  \left(  \nu \beta-\frac{\nu a_{1}-\mu
a_{2}}{\lambda}\right)  \right]  \qquad-\frac{1}{3}\left[  \left(  \tau
\gamma^{\ast}-\frac{\tau a_{2}^{\dag}-\nu a_{3}^{\dag}}{\lambda}\right)
\left(  \tau \gamma-\frac{\tau a_{2}-\nu a_{3}}{\lambda}\right)  \right]
\nonumber \\
&  \qquad \left.  -\frac{1}{3}\left[  \left(  \frac{\tau}{\nu}(\nu \beta^{\ast
}+\mu \gamma^{\ast})-\frac{\tau a_{1}^{\dag}-\mu a_{3}^{\dag}}{\lambda}\right)
\left(  \frac{\tau}{\nu}(\nu \beta+\mu \gamma)-\frac{\tau a_{1}-\mu a_{3}%
}{\lambda}\right)  \right]  \right \}  :\nonumber \\
&  =\left.  3\int_{-\infty}^{+\infty}\frac{dx}{\sqrt{6\pi}}:\exp \left[
-\frac{1}{3}\left(  \frac{3}{\sqrt{2}}x-\frac{\mu X_{1}+\nu X_{2}+\tau X_{3}%
}{\lambda}\right)  ^{2}\right]  :\right.  =1.
\end{align}

\subsection{The Conjugate State of $\left \vert \beta,\gamma,x\right \rangle $}

According to communication relationship between mechanic operator and quantum
state, once we known TNSCCE, we can derive its conjugate state.
Three-particle's total momentum is $P=\Sigma_{i=1}^{3}P_{i}$, $(P_{i}%
=(a_{i}-a_{i}^{\dag})/(\mathrm{i}\sqrt{2}))$, $P,$ $(\nu a_{1}-\mu a_{2})$,
and $(\tau a_{2}-\nu a_{3})$ are permutable with each other as well, we make
great efforts to find their common eigenvector with eigenvalues $\lambda
p/\sqrt{2}$, $\nu \sigma \lambda$ and $\tau \kappa \lambda$, expressed as
$\left \vert \sigma,\kappa,p\right \rangle $:%
\begin{align}
\left \vert \sigma,\kappa,p\right \rangle  &  =\exp \left \{  \left[  -\frac{3}%
{4}p^{2}-\frac{1}{6\nu}(\sigma^{\ast}\kappa+\sigma \kappa^{\ast})\mu \tau
^{2}-\frac{1}{6}|\kappa|^{2}\tau^{2}\left(  1+\frac{\mu^{2}}{\nu^{2}}\right)
-\frac{1}{6}|\sigma|^{2}\left(  \nu^{2}+\tau^{2}\right)  \right]  \right.
\nonumber \\
&  +\frac{1}{3\lambda}\left[  \left(  \sigma(\nu^{2}+\tau^{2})+\frac{\kappa
\mu \tau^{2}}{\nu}+3\mathrm{i}p\mu \right)  a_{1}^{\dag}+\left(  -\sigma \mu
\nu+\kappa \tau^{2}+3\mathrm{i}p\nu \right)  a_{2}^{\dag}+\left(  -\frac
{\kappa(\mu^{2}+\nu^{2})\tau}{\nu}-\sigma \mu \tau+3\mathrm{i}p\tau \right)
a_{3}^{\dag}\right] \nonumber \\
&  \left.  +\frac{1}{6\lambda^{2}}\left(  \mu a_{1}^{\dag}+\nu a_{2}^{\dag
}+\tau a_{3}^{\dag}\right)  ^{2}\right \}  \left \vert 000\right \rangle
\end{align}

The results after annihilation operators acting on $\left \vert \sigma
,\kappa,p\right \rangle $ respectively are
\begin{align}
a_{1}\left \vert \sigma,\kappa,p\right \rangle  &  =\frac{1}{3\lambda}\left[
\left(  \sigma \left(  v^{2}+\tau^{2}\right)  +\frac{k\mu \tau^{2}}{v}%
+3ip\mu \right)  +\frac{\mu}{\lambda}\left(  \mu a_{1}^{\dag}+va_{2}^{\dag
}+\tau a_{3}^{\dag}\right)  \right]  \left \vert \sigma,\kappa,p\right \rangle
.\nonumber \\
a_{2}\left \vert \sigma,\kappa,p\right \rangle  &  =\frac{1}{3\lambda}\left[
\left(  -\sigma \mu v+k\tau^{2}+3ipv\right)  +\frac{v}{\lambda}\left(  \mu
a_{1}^{\dag}+va_{2}^{\dag}+\tau a_{3}^{\dag}\right)  \right]  \left \vert
\sigma,\kappa,p\right \rangle ,\nonumber \\
a_{3}\left \vert \sigma,\kappa,p\right \rangle  &  =\frac{1}{3\lambda}\left[
\left(  -\frac{k\tau \left(  \mu^{2}+v^{2}\right)  }{v}-\sigma \mu \tau
+3ip\tau \right)  +\frac{\tau}{\lambda}\left(  \mu a_{1}^{\dag}+va_{2}^{\dag
}+\tau a_{3}^{\dag}\right)  \right]  \left \vert \sigma,\kappa,p\right \rangle ,
\label{16}%
\end{align}
After some combination of Eq.(\ref{16}), we get similar expressions as those
of TNSCCE
\begin{align*}
\frac{1}{3}\left(  \mu P_{1}+vP_{2}+\tau P_{3}\right)  \left \vert
\sigma,\kappa,p\right \rangle  &  =\frac{\lambda p}{\sqrt{2}}\left \vert
\sigma,\kappa,p\right \rangle ,\\
\left(  va_{1}-\mu a_{2}\right)  \left \vert \sigma,\kappa,p\right \rangle  &
=v\sigma \lambda \left \vert \sigma,\kappa,p\right \rangle ,\\
\left(  \tau a_{2}-va_{3}\right)  \left \vert \sigma,\kappa,p\right \rangle  &
=\tau k\lambda \left \vert \sigma,\kappa,p\right \rangle ,
\end{align*}
So we say that $\left \vert \sigma,\kappa,p\right \rangle $ is a tripartite
momentum non-symmetric coherent-entangled state.

\section{Generating The TNSCCE By Asymmetric Beamsplitter}

The tripartite coherent-entangled as we show upper, can be generated by
asymmetric BS operator. One function of BS it to generate entangled state
\cite{19}, and operator representation of BS operating on incident optic field
can be expressed by (with phase-free)\cite{20}
\begin{equation}
B_{ij}(\theta)=\exp \left[  -\theta \left(  a_{i}^{\dag}a_{j}-a_{i}a_{j}^{\dag
}\right)  \right]
\end{equation}

Letting the ideal single-mode maximal-squeezed state in mode 1, expressed by
$\left \vert x=0\right \rangle _{1}=exp\left(  -\frac{1}{2}a_{1}^{\dag2}\right)
\left \vert 0\right \rangle _{1}$, and the vacuum state $\left \vert
0\right \rangle _{2,3}$ in mode $2$, $3$ respectively enter the two input ports
of two sequential asymmetric BSs and get overlapped, we have%
\begin{align}
&  B_{23}(\varphi)B_{12}(\theta)\exp \left[  -\frac{1}{2}a_{1}^{\dag2}\right]
\left \vert 0\right \rangle _{1}\otimes \left \vert 0\right \rangle _{2}%
\otimes \left \vert 0\right \rangle _{3}\nonumber \\
&  =\exp \left[  -\frac{1}{2}\left(  a_{1}^{\dag}\cos(\theta)+a_{2}^{\dag}%
\sin(\theta)\cos(\varphi)+a_{3}^{\dag}\sin(\theta)\sin(\varphi)\right)
^{2}\right]  \left \vert 000\right \rangle . \label{17}%
\end{align}
When $\theta=\arccos(\frac{\sqrt{3}\mu}{3\lambda})$ and $\varphi=\arccos
(\frac{\nu}{\sqrt{\nu^{2}+\tau^{2}}})$, the state out of the two sequential
BSs in Eq.(\ref{17}) becomes $\exp \left[  -\frac{1}{6\lambda^{2}}\left(  \mu
a_{1}^{\dag}+\nu a_{2}^{\dag}+\tau a_{3}^{\dag}\right)  ^{2}\right]
\left \vert 000\right \rangle ,$ which is a three-mode squeezed state. Then
operating three sequential displacement operators $D_{1}(\epsilon_{1})$,
$D_{2}(\epsilon_{2})$, $D_{3}(\epsilon_{3})$ on three individual mode, where
$D_{i}(\epsilon)$ writes
\begin{equation}
D_{i}(\epsilon)=\exp(\epsilon a_{i}^{\dag}-\epsilon^{\ast}a_{i}),\qquad
i=1,2,3
\end{equation}
and the displacements $\epsilon_{1}$, $\epsilon_{2}$, $\epsilon_{3}$ are%
\begin{equation}
\epsilon_{1}=\frac{2\beta \upsilon \left(  v^{2}+\tau^{2}\right)  +2\gamma
\mu \tau^{2}+3x\mu \nu}{6\mu \lambda},\quad \epsilon_{2}=\frac{-2\beta \mu
v+2\gamma \tau^{2}+3xv}{6\lambda},\quad \varepsilon_{3}=\frac{-2\beta \mu \nu
\tau-2\gamma \tau \left(  \mu^{2}+v^{2}\right)  +3xv\tau}{6\nu \lambda}.
\label{18}%
\end{equation}
After these three sequential displacements, the ideal three-mode asymmetry
squeezed state will becomes%
\begin{align}
&  D_{1}(\epsilon_{1})D_{2}(\epsilon_{2})D_{3}(\epsilon_{3})\exp \left[
-\frac{1}{6\lambda^{2}}\left(  \mu a_{1}^{\dag}+\nu a_{2}^{\dag}+\tau
a_{3}^{\dag}\right)  ^{2}\right]  \left \vert 000\right \rangle \nonumber \\
&  =\exp \left \{  -\frac{\epsilon_{1}\epsilon_{1}^{\ast}+\epsilon_{2}%
\epsilon_{2}^{\ast}+\epsilon_{3}\epsilon_{3}^{\ast}}{2}+\epsilon_{1}%
a_{1}^{\dag}+\epsilon_{2}a_{2}^{\dag}+\epsilon_{3}a_{3}^{\dag}\right.
\nonumber \\
&  \left.  -\frac{1}{6\lambda^{2}}\left(  \mu(a_{1}^{\dag}-\epsilon_{1}^{\ast
})+\nu(a_{2}^{\dag}-\epsilon_{2}^{\ast})+\tau(a_{3}^{\dag}-\epsilon_{3}^{\ast
})\right)  ^{2}\right \}  \left \vert 000\right \rangle
\end{align}

Experimentally, we can achieve these displacements (eg. $D_{1}(\epsilon_{1}%
)$), by reflecting the light field\newline$\exp \left[  -\frac{1}{6\lambda^{2}%
}\left(  \mu a_{1}^{\dag}+\nu a_{2}^{\dag}+\tau a_{3}^{\dag}\right)
^{2}\right]  \left \vert 000\right \rangle $ from a partially reflecting mirror
(say $99\%$ reflection and $1\%$ transmission) and adding through the mirror a
field that has been phase and amplitude modulated according to the values
$\mu$, $\nu$, $\tau$, and $\beta$, $\gamma$, $x$. Thus the Tripartite
Coherent-Entangled state $\left \vert \beta,\gamma,x\right \rangle $ can be implemented.

\section{Some Applications of $\left \vert \beta,\gamma,x\right \rangle $}

\subsection{Three-mode Wigner Operator derived from $\left \vert \beta
,\gamma,x\right \rangle $}

In atomic and quantum optic area, Wigner distribution as the quasiclassical
distribution \cite{21,22} well represent the nonclasscial properties of
quantum state throught its partial negativity in quadrature phase.One basic
way to obtain Wigner distribution is to trace production between matrix
density and Wigner operator \cite{23}. Analogue to single mode Wigner
function, and basing on completeness and orthogonality of TNSCE, we now
introduce the following ket-bra integration
\begin{equation}
\iint \frac{d^{2}\beta}{\pi}\frac{d^{2}\gamma}{\pi}\int_{-\infty}^{+\infty
}\frac{du}{2\pi \sqrt{6\pi}}e^{3\mathrm{i}pu/2}\left \vert \beta,\gamma
,x+\frac{u}{2}\right \rangle \left \langle \beta,\gamma,x-\frac{u}{2}\right \vert
\equiv \triangle(p,x)
\end{equation}
Considering the explicit definition of $\left \vert \beta,\gamma,x\right \rangle
$ in Eq.(\ref{8}) and the IWOP technique, we can directly calculate out
\begin{equation}
\triangle(p,x)=\frac{1}{\pi \tau^{2}\lambda^{2}}:\exp \left[  -3\left(  \frac
{x}{\sqrt{2}}-\frac{\mu X_{1}+\nu X_{2}+\tau X_{3}}{3\lambda}\right)
^{2}-3\left(  \frac{p}{\sqrt{2}}-\frac{\mu P_{1}+\nu P_{2}+\tau P_{3}%
}{3\lambda}\right)  ^{2}\right]  : \label{19}%
\end{equation}

which is a generalization of the normally ordered form of the usual Wigner
operator. We may integrate $\Delta(p,x)$ out of $\emph{x,p},$ respectively,
e.g.
\begin{align}
\int_{-\infty}^{\infty}dx\Delta \left(  p,x\right)   &  =\sqrt{\frac{2}{3\pi}%
}\frac{1}{\tau^{2}\lambda^{2}}:\exp \left[  -3\left(  \frac{1}{\sqrt{2}}%
p-\frac{\mu P_{1}+vP_{2}+\tau P_{3}}{3\lambda}\right)  ^{2}\right]
:\nonumber \\
&  =\sqrt{\frac{1}{6\pi}}%
{\displaystyle \iint}
\frac{d^{2}\sigma d^{2}k}{\pi^{2}}\left \vert \sigma,k,p\right \rangle
\left \langle \sigma,k,p\right \vert ,
\end{align}

\begin{align}
\int_{-\infty}^{\infty}dp\Delta \left(  p,x\right)   &  =\sqrt{\frac{2}{3\pi}%
}\frac{1}{\tau^{2}\lambda^{2}}:\exp \left[  -3\left(  \frac{1}{\sqrt{2}}%
x-\frac{\mu X_{1}+vX_{2}+\tau X_{3}}{3\lambda}\right)  ^{2}\right]
:\nonumber \\
&  =\sqrt{\frac{1}{6\pi}}%
{\displaystyle \iint}
\frac{d^{2}\beta d^{2}\gamma}{\pi^{2}}\left \vert \beta,\gamma,x\right \rangle
\left \langle \beta,\gamma,x\right \vert ,
\end{align}
Following Wigner's original idea of setting up a function in $x$-$p$ phase
whose marginal distribution is the probability of finding a particle in
coordinate space and momentum space, respectively, we can immediately judge
that the operator $\Delta(p,x)$ in Eq.(\ref{19}) is just a marginal
distributional Wigner operator. Then the marginal distribution in the
$p$-direction and its conjugate marginal distributions in the $x$-direction
are
\begin{align}
\int_{-\infty}^{\infty}dx\left \langle \psi \right \vert \Delta \left(
p,x\right)  \left \vert \psi \right \rangle  &  =\sqrt{\frac{2}{3\pi}}\frac
{1}{\tau^{2}\lambda^{2}}\left \langle \psi \right \vert :\exp \left[  -3\left(
\frac{1}{\sqrt{2}}p-\frac{\mu P_{1}+vP_{2}+\tau P_{3}}{3\lambda}\right)
^{2}\right]  :\left \vert \psi \right \rangle \nonumber \\
&  =\sqrt{\frac{1}{6\pi}}%
{\displaystyle \iint}
\frac{d^{2}\sigma d^{2}k}{\pi^{2}}\left \vert \langle \psi \left \vert
\sigma,k,p\right \rangle \right \vert ^{2},
\end{align}
and%
\begin{align}
\int_{-\infty}^{\infty}dp\left \langle \psi \right \vert \Delta \left(
p,x\right)  \left \vert \psi \right \rangle  &  =\sqrt{\frac{2}{3\pi}}\frac
{1}{\tau^{2}\lambda^{2}}\left \langle \psi \right \vert :\exp \left[  -3\left(
\frac{1}{\sqrt{2}}x-\frac{\mu X_{1}+vX_{2}+\tau X_{3}}{3\lambda}\right)
^{2}\right]  :\left \vert \psi \right \rangle \nonumber \\
&  =\sqrt{\frac{1}{6\pi}}%
{\displaystyle \iint}
\frac{d^{2}\beta d^{2}\gamma}{\pi^{2}}\left \vert \langle \psi \left \vert
\beta,\gamma,x\right \rangle \right \vert ^{2},
\end{align}
correspondingly. Furthermore, we should note that in this case the classical
$x$-$p$ phase space corresponds to the operators $X$ and $P$ respectively.

\subsection{Three-mode Squeezing Operator}

One important application of IWOP technique is to construct squeezed operator
no matter how complex the quantum states is in continuous variable quadrature
space \cite{26,27}. In a similar way, We take a classical transformation
$x\rightarrow \frac{x}{l}$ in $\left \vert \beta,\gamma,x\right \rangle $ to
build a ket-bra integral,
\begin{equation}
S(l)=\iint \frac{1}{\pi^{2}}d^{2}\beta d^{2}\gamma \int_{-\infty}^{+\infty}%
\frac{dx}{\sqrt{6l\pi}}\left \vert \beta,\gamma,x/l\right \rangle \left \langle
\beta,\gamma,x\right \vert \label{20}%
\end{equation}
Using the IWOP technique, we can directly perform the integration in
Eq.(\ref{20}) to obtain
\begin{align}
S(l)  &  =\frac{1}{\tau^{2}\lambda^{2}}sech^{1/2}(\lambda)\exp \left \{
-\frac{1}{6}(\mu a_{1}^{\dag}+\nu a_{2}^{\dag}+\tau a_{3}^{\dag})^{2}%
\tanh \lambda \right \} \nonumber \\
&  :\exp \left \{  \frac{1}{3}(sech\lambda-1)(\mu a_{1}^{\dag}+\nu a_{2}^{\dag
}+\tau a_{3}^{\dag})(\mu a_{1}+\nu a_{2}+\tau a_{3})\right \}  :\nonumber \\
&  \exp \left \{  \frac{1}{6}(\mu a_{1}+\nu a_{2}+\tau a_{3})^{2}\tanh
\lambda \right \}  \label{21}%
\end{align}
which is a three-mode squeezing operator. where $l=\exp(\lambda)$,
$sech{}\lambda=2l/(1+l^{2})$ and $\tanh{}\lambda=(l^{2}-1)/(1+l^{2})$. To make
this squeezing more compact, we introduce the notation $R=\frac{\mu
a_{1}^{\dag}+\nu a_{2}^{\dag}+\tau a_{3}^{\dag}}{\sqrt{3}\lambda}$, and using
the following formula $:\exp[(e^{\zeta}-1)a^{\dag}a]:=\exp(\zeta a^{\dag}a)$,
we can rewrite the Eq.(\ref{21}) as
\begin{align}
S(l)  &  =\frac{1}{\tau^{2}\lambda^{2}}sech^{1/2}(\lambda)\exp \left \{
-\frac{1}{2}R^{\dag2}\tanh \lambda \right \}  :\exp \left \{  (sech\lambda
-1)R^{\dag}R\right \}  :\exp \left \{  \frac{1}{2}R^{2}\tanh \lambda \right \}
\nonumber \\
&  =\frac{1}{\tau^{2}\lambda^{2}}sech^{1/2}(\lambda)\exp \left \{  -\frac{1}%
{2}R^{\dag2}\tanh \lambda \right \}  \exp \left \{  R^{\dag}R\ln sech\lambda
\right \}  \exp \left \{  \frac{1}{2}R^{2}\tanh \lambda \right \}  .
\end{align}
We find that $R^{2},R^{\dag2},$ $\left(  R^{\dag}R+\frac{1}{2}\right)  $
compose a $SU(1,1)$ Lie algebra as
\begin{equation}
\lbrack R^{2},R^{\dag2}]=2\left(  R^{\dag}R+\frac{1}{2}\right)  ,\qquad \left[
\left(  R^{\dag}R+\frac{1}{2}\right)  ,R^{2}\right]  =-R^{2},\qquad \left[
\left(  R^{\dag}R+\frac{1}{2}\right)  ,R^{\dag2}\right]  =R^{\dag2}.
\end{equation}
Correspondingly, the three-mode squeezed vacuum state is
\begin{equation}
S(l)\left \vert 000\right \rangle =sech^{1/2}(\lambda)\frac{1}{\tau^{2}%
\lambda^{2}}\exp \left \{  -\frac{1}{6}(\mu a_{1}^{\dag}+\nu a_{2}^{\dag}+\tau
a_{3}^{\dag})^{2}\tanh \lambda \right \}  \left \vert 000\right \rangle .
\end{equation}

\section{New Multipartite Coherent-Entangled State and Its Generation}

In order to introduce the new multipartite CES and its generation, we should
carefully study the generation of the bipartite CES and tripartite CES in more
detail. For the bipartite CES in Eq.(\ref{4}), similarly to the tripartite
CES, can be generated. Letting the ideal single-mode maximal-squeezed state in
mode $1$ and the vacuum state in mode 2 respectively enter an asymmetry BS to
get overlapped and then incident into two sequential displacement operator
$D_{1}(\epsilon_{1}=\frac{\mu x+\nu^{2}\alpha}{\lambda})$ and $D_{2}%
(\epsilon_{2}=\frac{\nu x+\mu \nu \alpha}{\lambda})$, we can obtain $\left \vert
\alpha,x\right \rangle $ in Eq.(\ref{4}).
\begin{equation}
\left \vert \alpha,x\right \rangle =D_{1}(\epsilon_{1})D_{2}(\epsilon_{2}%
)B_{12}(\theta)\left \vert x=0\right \rangle _{1}\otimes \left \vert
0\right \rangle _{2}.
\end{equation}
Introducing $\delta_{i}\left(  i=1,2\right)  $ which satisfy:
\begin{equation}
\alpha=\delta_{1}-\frac{\mu}{v}\delta_{2},
\end{equation}
so we can rewrite the displacements of the two displacement operators%
\begin{equation}
\epsilon_{1}=\frac{\delta_{1}v^{2}-\delta_{2}\mu \nu+\mu x}{\lambda}%
,\qquad \epsilon_{2}=\frac{-\delta_{1}\mu \nu+\delta_{2}v^{2}+vx}{\lambda}.
\label{22}%
\end{equation}
For the new tripartite CES, we also introduce the $\delta_{i}$ $\left(
i=1,2,3\right)  $for simplifity the displacements
\begin{equation}
\beta=\delta_{1}-\frac{\mu}{v}\delta_{2},\qquad \gamma=\delta_{2}-\frac{v}%
{\tau}\delta_{3},
\end{equation}
then substitute $\beta$, $\gamma$ to Eq.(\ref{18}), the displacements will be
\begin{align}
\epsilon_{1}  &  =\frac{1}{3\lambda}\left[  \delta_{1}\left(  v^{2}+\tau
^{2}\right)  -\delta_{2}\mu v-\delta_{3}\mu \tau+\frac{3}{2}\mu x\right]
,\nonumber \\
\epsilon_{2}  &  =\frac{1}{3\lambda}\left[  -\delta_{1}\mu v+\delta_{2}\left(
v^{2}+\tau^{2}\right)  -\delta_{3}v\tau+\frac{3}{2}vx\right]  ,\nonumber \\
\epsilon_{2}  &  =\frac{1}{3\lambda}\left[  -\delta_{1}\mu \tau-\delta_{2}%
v\tau+\delta_{3}\left(  \mu^{2}+\nu^{2}\right)  +\frac{3}{2}\tau x\right]  .
\label{23}%
\end{align}
For the multi-partite CES, which can be constructed by letting $\left \vert
x=0\right \rangle _{1}\otimes \left \vert 0\cdots0\right \rangle _{2...N}$
incident into a sequential of $N-1$ asymmetry BSs, we have%
\begin{align}
&  B_{N,N-1}(\theta_{N-1})B_{N-1,N-2}(\theta_{N-2})\cdots B_{2,1}(\theta
_{1})\left \vert x=0\right \rangle _{1}\otimes \left \vert 0\cdots0\right \rangle
_{2...N}\nonumber \\
&  =\exp \left[  -\frac{1}{2}\left(  a_{1}^{\dag}\cos(\theta_{1})+\cdots
+\left(  a_{i}^{\dag}\sin(\theta_{1})\sin(\theta_{2})\cdots \sin(\theta
_{i-1})\cos(\theta_{i})\right)  \right.  \right.  \, \nonumber \\
&  \,+\left.  \left.  \cdots+a_{N}^{\dag}\sin(\theta_{1})\sin(\theta
_{2})\cdots \sin(\theta_{N-1})\right)  ^{2}\right]  \left \vert 0\cdots
0\right \rangle _{2...N}.
\end{align}
Noting$\frac{\mu_{i}}{\sqrt{N}\lambda}=\sin(\theta_{1})\sin(\theta_{2}%
)\cdots \sin(\theta_{i-1})\cos(\theta_{i})$, the state is
\begin{equation}
\left \vert \mu_{1},\mu_{2},\cdots,\mu_{N}\right \rangle =\exp \left[  -\frac
{1}{2N\lambda^{2}}\left(  \sum_{i=1}^{N}\mu_{i}a_{i}^{\dag}\right)
^{2}\right]  \left \vert 0\cdots0\right \rangle _{1\ldots N}.
\end{equation}
We can get the new Multi-partite CES $\left \vert \beta_{1},\beta_{2}%
,\cdots,\beta_{N-1},x\right \rangle $ by operating $N-1$ displacement on
$\left \vert \mu_{1},\mu_{2},\cdots,\mu_{N}\right \rangle $. That is $\left \vert
\beta_{1},\beta_{2},\cdots,\beta_{N-1},x\right \rangle =\prod_{i=1}^{N-1}%
D_{i}(\epsilon_{i})\left \vert \mu_{1},\mu_{2},\cdots,\mu_{N}\right \rangle $
and the new CES satisfies the following $N$ eigen-equations.
\begin{align}
\frac{1}{N}\sum_{i=1}^{N}\mu_{i}X_{i}\left \vert \beta_{1},\beta_{2}%
,\cdots,\beta_{N-1},x\right \rangle  &  =\frac{\lambda x}{\sqrt{2}}\left \vert
\beta_{1},\beta_{2},\cdots,\beta_{N-1},x\right \rangle ,\nonumber \\
\left(  \mu_{i+1}a_{i}-\mu_{i}a_{i+1}\right)  \left \vert \beta_{1},\beta
_{2},\cdots,\beta_{N-1},x\right \rangle  &  =\mu_{i+1}\beta_{i}\left \vert
\beta_{1},\beta_{2},\cdots,\beta_{N-1},x\right \rangle ,\qquad \nonumber \\
i  &  =1,2,...,N-1.
\end{align}
In order to write down the explicit displacements, we also introduce
$\delta_{i}$
\begin{equation}
\mu_{i+1}\beta_{i}=\mu_{i+1}\delta_{i}-\mu_{i}\delta_{i+1},\qquad
i=1,2,\cdots,N-1
\end{equation}
Referenced to Eq.(\ref{22}) and (\ref{23}), the displacements of the $N-1$
displacement operators $D_{i}(\epsilon_{i})$, $i=1,2,\ldots,N$ will be
\begin{equation}
\epsilon_{i}=\frac{1}{N\lambda}\left(  \delta_{i}\sum_{j,j\neq i}^{N}\left(
\mu_{j}^{2}\right)  -\mu_{i}\sum_{j,j\neq i}^{N}\left(  \delta_{j}\mu
_{j}\right)  +\frac{N}{2}\mu_{i}x\right)
\end{equation}
Similar argument goes as the generation of New Tripartite CES will show that
the new Multi-partite CES $\left \vert \beta_{1},\beta_{2},\cdots,\beta
_{N-1},x\right \rangle $ can also be implemented.

\section{Conclusion}

In summary, we have brought out the ways to construct TNSCCE just contrary the
traditional method and check it correckness.We then analyze the new state's
properties, i.e. the completeness relation and partly orthogonality. And a
simple experimental protocol to produce TNSCCE using an asymmetric
beamsplitter was also proposed. It provide a new way to predict new tripartite
squeezed operator. And basing on this way to construct TNSCCE, we predict
multi-partite non-symmetric entangled state can be gotten.\newline

\end{document}